\documentstyle[12pt,epsf]{article}

\def\mydate{February 17, 1995}

\normalsize

\newcounter{sxn}

\newcounter{axn}

\date{}

\newdimen\mybaselineskip
\newcommand{\beeq}{\begin{equation}}
\newcommand{\eneq}{\end{equation}}
\newcommand{\beqn}{\begin{eqnarray}}
\newcommand{\eeqn}{\end{eqnarray}}

\def\mybig{\displaystyle \strut }
\def\mbig{\displaystyle }
\def\d{\partial}
\def\la{\raise.16ex\hbox{$\langle$} \, }
\def\ra{\, \raise.16ex\hbox{$\rangle$} }
\def\lla{\raise.16ex\hbox{$\langle$} \kern-.25em\raise.16ex\hbox{$\langle$}\, }
\def\rra{\,\raise.16ex\hbox{$\rangle$}
     \kern-.25em\raise.16ex\hbox{$\rangle$}}
\def\go{\rightarrow}

\def\next{{~~~,~~~}}
\def\onehalf{ \hbox{${1\over 2}$} }
\def\psibar{ \psi \kern-.65em\raise.6em\hbox{$-$} }
\def\chibar{ \chi \kern-.65em\raise.5em\hbox{$-$} }
\def\mbar{ m \kern-.78em\raise.4em\hbox{$-$}\lower.4em\hbox{} }
\def\delbar{{\bar\delta}}
\def\Bbar{ B \kern-.73em\raise.6em\hbox{$-$}\hbox{} }

\def\ZZ{ \hbox{\tenss Z} \kern-.4em \hbox{\tenss Z} }

\def\ep{\epsilon}

\def\wil{ \Theta_{\rm w} }

\def\vphi{ {\varphi} }

\def\EM{{\rm EM}}

\def\mass{{\rm mass}}
\def\osc{{\rm osci}}

\def\vPhi{{\Phi_{\rm vac}}}

\def\sign{\,{\rm sign}\,}

\begin{document}

\bibliographystyle{unsrt}
\footskip 1.0cm
\thispagestyle{empty}

{\baselineskip=10pt \parindent=0pt \small
\mydate \hfill \hbox{\vtop{\hsize=3.2cm UMN-TH-1324/95 \\ AZPH-TH/95-1\\
           UT-698\\}}
\vspace{20mm}
}

\centerline {\LARGE\bf  The Massive Multi-flavor Schwinger Model}
\vspace*{20mm}
\centerline {\large  J. E. Hetrick,$^{1}$ Y. Hosotani,$^{2}$ and
       S. Iso$^{3}$}

\vspace*{5mm}
\centerline {\small\it $^{1}$Department of Physics, University of Arizona}
\centerline {\small\it Tucson, Arizona 85721, U.S.A.}
\centerline {\small\sf hetrick@physics.arizona.edu}
\bigskip
\centerline {\small\it $^{2}$School of Physics and Astronomy, University
       of  Minnesota}
\centerline {\small\it Minneapolis, Minnesota 55455, U.S.A.}
\centerline {\small\sf yutaka@mnhep.hep.umn.edu}
\bigskip
\centerline {\small\it $^{3}$Department of Physics, University of Tokyo}
\centerline {\small\it Tokyo 113, Japan}
\centerline {\small\sf iso@danjuro.phys.s.u-tokyo.ac.jp}
\vspace*{15mm}
%\baselinestretch{2.0}
\normalsize

\begin{abstract}
\baselineskip=17pt
QED with $N$ species of massive fermions on a circle of
circumference $L$ is analyzed by bosonization.
The problem is
reduced to the quantum mechanics of the $2N$ fermionic and one
gauge field zero modes on the circle, with nontrivial interactions
induced by the chiral anomaly and fermions masses. The solution is given
for $N=2$ and fermion masses ($m$) much smaller than the mass of the
$U(1)$ boson  with mass $\mu=\sqrt{2e^2/\pi}$
when all fermions satisfy the same boundary conditions.
We show that the two limits $m \go 0$ and $L \go \infty$ fail to
commute and that the behavior of the theory critically
depends on the value of  $mL|\cos\onehalf\theta|$
where $\theta$ is the vacuum angle parameter.
When the volume is large $\mu L \gg 1$, the fermion condensate
$\la \psibar \psi \ra$
is  $-\big( e^{4\gamma} m\mu^2 \cos^4\onehalf\theta/4\pi^3)^{1/3} $
or $-2e^\gamma m\mu L \cos^2 \onehalf\theta /\pi^2$
for $mL(\mu L)^{1/2} |\cos\onehalf\theta| \gg 1$ or  $\ll 1$,
respectively.    Its correlation function decays algebraically with a
critical exponent $\eta=1$ when $m\cos\onehalf\theta=0$.
\end{abstract}

\newpage

\baselineskip=22pt

%%%%%%%%%%%%%%%%%%%%%%%%%%%%%%%%%%%%%%%%%%%%%%%%%%%%%%%%%%%%%%%%%

The Schwinger model, QED in two dimensions, with $N\ge2$ species of
fermions is distinctly different from that with one flavor
\cite{Schwinger}$-$\cite{Affleck1}. For example,
Affleck has shown that in the massless fermion case, one
massive boson of mass $\mu=\sqrt{Ne^2/\pi}$ and $N-1$ massless bosons
appear, however there is
no long range order ($\la \psibar \psi\ra=0$) in accordance with
Coleman's theorem in a 2-d Lorentz invariant theory
\cite{Affleck1,Coleman0} and correlators of $\psibar\psi$ show
algebraic decay at large distances.  Hence the rich vacuum structure
of the multi-flavor Schwinger model carries many similarities to
4-dimensional QCD where we are interested in understanding how the
effects of quark masses modify the vacuum structure, meson masses,
mixing, and the pattern of chiral symmetry breaking.

Years ago Coleman showed \cite{Coleman} that in the presence of small
fermion masses $m \ll \mu$ in the $N=2$ model, the second boson mass
has a fractional power dependence on $m$ and the vacuum angle
$\theta$: $\mu_2 \propto (m |\cos\onehalf\theta|)^{2/3}$. This
singular dependence poses an intriguing puzzle: how can one get
non-analytic dependence in the $m\go 0$ limit where one would expect
the validity of a perturbation theory in mass?

Thus there has been growing interest in the Schwinger model,
especially when defined on a compact manifold such as a circle or
closed interval (a bag)\cite{Manton}-\cite{Saradzhev}. Besides reproducing
results in Minkowski spacetime in the infinite volume limit,
analyzing the model on a circle has the advantage of being free from
infrared divergence and well-defined at every stage of manipulation.
Furthermore analytic solutions of the multi-flavor model are
extremely useful for comparison with lattice simulations where
several flavors are inherent \cite{Dilger}\cite{Bock}\cite{MPR}-\cite{BK}.

In this paper we solve the Schwinger model with two massive fermions
on a circle of circumference $L$.  We find that the theory
sensitively depends on the dimensionless parameter
$mL\cos\onehalf\theta$.  In particular, the $m\go 0$ and $L\go \infty$
limits do not commute with each other.  This is  to be contrasted
to the situation in a model with just one fermion, in which a small
fermion mass does not alter the structure of the model except for
necessitating a $\theta$ vacuum.

In the $SU(2)$ symmetric two flavor case ($m_1 = m_2$), we show
that in the large volume $\mu L \gg 1$ limit, the light boson mass
$\mu_2 \propto (m |\cos\onehalf\theta|)^{2/3}$ ~for~ $mL(\mu L)^{1/2}
|\cos\onehalf\theta| \gg 1$, while it is $m |\cos\onehalf\theta|$ ~for~
$mL(\mu L)^{1/2} |\cos\onehalf\theta| \ll 1$.
In other words physical quantities behave
smoothly in the $m\go 0$ or $\theta\go \pm \pi$ limit.

Several authors have given exact solutions for the $N\ge 2$ model with
massless fermions on various manifolds.\cite{Seiler,Joos2,Wipf} Yet
the importance of the parameter $mL|\cos\onehalf\theta|$ has not been
stressed in the literature. We adopt the method of abelian
bosonization on a circle, generalizing the analysis of the $N$=1 case
in ref.\ \cite{HH}.  With $N$ fermions the problem is eventually
reduced to quantum mechanics for the $2N+1$ ``zero modes'' on the
circle, with nontrivial interactions induced by the chiral anomaly and
fermions masses. Further reduction is achieved for $m$$\ll$$\mu$. We
find that the wave function of the vacuum
sensitively depends on fermion masses for $N \ge 2$.

The model  is given by
\beeq
{\cal L} = - \hbox{$1\over 4$} \, F_{\mu\nu} F^{\mu\nu} +
\sum_{a=1}^N \psibar_a \Big\{ \gamma^\mu (i \d_\mu - e A_\mu) -
  m_a  \Big\}
   \psi_a  ~.
\eneq
Fermion fields obey boundary conditions
\beeq
\psi_a(t,x+L)
      = -\, e^{2\pi i \alpha_a } \, \psi_a(t,x)
  ~. \label{BC1}
\eneq
We suppose that gauge fields are periodic.   Note that with
$\alpha_a=0$ the system is mathematically equivalent to the Schwinger
model on a line at finite temperature $T=L^{-1}$.

It is possible to choose a gauge in which
\beeq
A_0(t,x) = - \int_0^L dy \, G(x-y) \, j_\EM^0(t,y)  \next
A_1(t,x) = {\Theta_W(t)\over eL}
\eneq
where $j_\EM^0 = e \sum_a \psi_a^\dagger \psi_a$ and $G(x)$ is the
periodic Green's function on a circle satisfying
$ G''(x) = \delta_L(x) - L^{-1}$.  $\Theta_W(t)$ is the non-integrable phase of
the Wilson line integral around the circle, and is a physical degree of
freedom.  In this Coulomb gauge the Hamiltonian is
\beqn
H &=& {e^2L\over 2} \, \Pi_\Theta^2 +
\int_0^L dx \, \sum_a \psibar_a \Big\{
\gamma^1 \Big(-i\d_1 + {\Theta_W\over L} \Big) +  m_a \Big\} \psi_a  \cr
&&\hskip 2cm - {1\over 2} \int_0^L dxdy ~ j_\EM^0(x) G(x-y) j_\EM^0(y) ~~.
    \label{Hamiltonian}
\eeqn
$\Pi_\Theta$ is the  momentum conjugate to $\Theta_W$:
$\Pi_\Theta=\dot \Theta_W / e^2L$, and
the anti-symmetrization of fermion operators is understood.

Fermions are bosonized first in the interaction picture defined by free
massless fermions. We take $\gamma^\mu = (\sigma_1, i\sigma_2)$ so that
$\psi_a^t=(\psi^a_+,\psi^a_-)$ satisfies $(\d_0 \pm \d_1) \psi^a_\pm =0$,
and introduce $2N$ sets of bosonic variables $\{q,p, c_n^{}, c_n^\dagger \}$
satisfying
\beqn
&&[q^a_\pm, p^b_\pm] = i \, \delta^{ab}  ~~~,~~~
[c^a_{\pm,n}, c^{b,\dagger}_{\pm,m}] = \delta^{ab} \delta_{nm} \cr
&&\phi^a_\pm (t,x) = \sum_{n=1}^\infty {1\over \sqrt{n}} \,
  \Big\{ c^a_{\pm,n} \, e^{- 2\pi in(t \pm x)/L} + {\rm h.c.} \Big\}
\eeqn
In terms of these variables the $\psi^a_\pm(x)$'s are represented as
\beeq
\psi^a_\pm (t,x) = {1\over \sqrt{L}} \, C^a_\pm \,
 e^{\pm i \{ q^a_\pm + 2\pi p^a_\pm (t \pm x)/L \} }
  :\, e^{\pm i\phi^a_\pm (t,x) } \, :   \label{bosonize}
\eneq
where the Klein factors $C^a_\pm$ are defined by
\beeq
C^a_+ =\exp \Big\{ i\pi \sum_{b=1}^{a-1} ( p^b_+ + p^b_- - 2 \alpha_b)
   \Big\}  ~~~,~~~
C^a_- = \exp \Big\{ i\pi \sum_{b=1}^{a} ( p^b_+ - p^b_-) \Big\}
\eneq
and $:~:$ indicates normal ordering with respect to
$(c_n^{},c_n^\dagger)$.  It is straightforward to show that at equal time
\beeq
\{ \psi^a_\pm(x), \psi^b_\pm(y)^\dagger \}
= \delta^{ab}  ~ e^{i\pi (x-y)/L} \cdot e^{2\pi i p^a_\pm (x-y)/L}
{}~ \delta_L(x-y)    \label{AntiCommutator}
\eneq
and that all other anti-commutators vanish.
Further
\beeq
\psi^a_\pm (t,x+L) = - e^{2\pi i p^a_\pm} \, \psi^a_\pm (t,x) =
-  \psi^a_\pm (t,x) \, e^{2\pi i p^a_\pm}
\eneq
so that the boundary  condition (\ref{BC1}) is guaranteed if
\beeq
e^{2\pi i p^a_\pm} ~ | \, {\rm phys} \ra = e^{2\pi i \alpha_a} ~
|\, {\rm phys} \ra ~~.   \label{PhysState}
\eneq
With this physical state condition the anticommutation relations
(\ref{AntiCommutator}) are consistent with the boundary  condition
(\ref{BC1}).  We note that
$(C^a_\pm {\rm ~or~} {C^a_\pm }^\dagger)
 \, | \, {\rm phys} \ra =   | \, {\rm phys} \ra $.

The bosonized currents and Hamiltonian are easily deduced by direct
substitution of (\ref{bosonize}).
The flavor diagonal  currents
$j_a^\mu=\psibar_a \gamma^\mu \psi_a^{}$ are
\beqn
j^0_a &=& {-p^a_+ + p^a_- \over L} - {1\over 2\pi} \, \d_x (\phi^a_+ +
\phi^a_-) \cr
\noalign{\kern 6pt}
j^1_a &=& {+p^a_+ + p^a_- \over L}
+ {1\over 2\pi} \, \d_t (\phi^a_+ + \phi^a_-)  +  {\Theta_W \over \pi L} ~.
\eeqn
It will prove convenient to rotate to a new basis in the flavor space.
Introduce  an orthogonal $N$-by-$N$ matrix ${v^\alpha}_a $,  where
${v^1}_a =1/\sqrt{N}$; $N$ new fields defined by
$\chi^\alpha_\pm = (4\pi)^{-1/2} \,{v^\alpha}_a  \phi^a_\pm$; and let
$\chi_\alpha = \chi^\alpha_+ + \chi^\alpha_-$.
Note that $j^\mu_\EM = e \sum_a j^\mu_a
=  - \mu \ep^{\mu\nu} \d_\nu \chi_1 + \cdots$ where $\mu^2 = N e^2/\pi$.
The $\chi_1^{}$ field  represents the charged part.

Then the  Hamiltonian in the Schr\"odinger picture becomes
\beqn
H &=& H_0 + H_\osc  + H_\mass + ({\rm constant}) \cr
\noalign{\kern 6pt}
H_0 &=&  {e^2 L\Pi_\Theta^2\over 2}
+ {\pi\over 2L} \sum_{a=1}^N \Big\{ (p^a_+-p^a_-)^2
+ (p^a_+ + p^a_- + {\Theta_W\over \pi} )^2 \Big\} \cr
H_\osc &=& \int_0^L dx \, {1\over 2}
\bigg\{ ~ N_\mu[ {\Pi_1}^2 + {\chi_1'}^2   + \mu^2 \,  \chi_1^2]
 + \sum_{\alpha=2}^N N_0[\Pi_\alpha^2 + {\chi_\alpha'}^2 ] ~ \bigg\} \cr
H_{\rm mass} &=& \int_0^Ldx \, \big\{ m_1 M_{11}
                              +  m_2 M_{22}+ ~{\rm h.c.}~ \big\} ~~,
    \hskip .5cm M_{ab}  = \psi^{a \dagger}_- \psi^b_+ .
\label{Hamiltonian2}
\eeqn
$\Pi_\alpha= \dot\chi_\alpha$ is the conjugate momentum to
$\chi_\alpha$.  $N_\mu[\cdots]$ denotes normal ordering in the
Schr\"odinger picture with respect to a mass $\mu$.
%$H_\mass$ represents the bare fermion mass term.
In the massless fermion limit
we have one massive boson $\chi_1$
and $(N-1)$ massless bosons $\chi_\alpha$ $(\alpha=2
\sim N)$ irrespective of boundary conditions of fermions.
Physical states must satisfy (\ref{BC1}) and $Q^\EM |{\rm phys}\ra =0$
where $Q^\EM= e \sum_a (-p^a_+ + p^a_-)$.

In the absence of $H_{\rm mass}$ , the $2N+1$ zero modes
($\Theta_W, q^a_\pm$) and the oscillatory modes ($\chi_\alpha$)
decouple.  Thus our strategy is as follows: we first determine all
matrix elements of the total Hamiltonian $H$ in the basis spanned by
eigenstates of $H_0+H_\osc$, and then diagonalize it in the case
$m_a/\mu \ll 1$.  As we shall see below, in a large volume the fermion
mass term $H_\mass$ cannot be treated as a small perturbation to $H_0
+ H_\osc$.  On the contrary it dominates over $H_0$ and for $N \ge 2$,
$H_\mass$ completely alters the structure of the vacuum.

Let us restrict ourselves henceforth to the case $N=2$  and $\alpha_a
=\alpha$.  $H_\osc$ describes free fields.  The boundary condition
(\ref{PhysState}) implies that $q^a_\pm$'s are angular variables so that wave
functions can be expanded in a Fourier series in each zero mode:
$e^{in^a_\pm q^a_\pm}$.  The relevant eigenstates
of
$H_0$ are
\beeq
\Phi_s^{(n,r)} = {1\over (2\pi)^2} \,
u_s(\Theta_W + 2\pi n + \pi r + 2\pi \alpha)
{}~ e^{i(n+\alpha)(q^1_++q^1_-) + i(n+r+\alpha)(q^2_++q^2_-)}
\eneq
with energy
\beeq
E_s^{(n,r)} =  ~ \mu \Big( s +{1\over 2} \Big) + {\pi r^2\over L}
   \label{WaveFun}
\eneq
where ($n,r$) are integers.  $u_s(x)$ ($s=0,1,2, \cdots$) is the
$s$-th eigenfunction in a harmonic oscillator problem. In particular
$u_0(x)=( 2/ \mu L \pi^2)^{1/4} \, e^{-x^2/\pi\mu L}$.  One need
consider only states with $n^a_+=n^a_-$, since $n^a_+ \not= n^a_-$
gives a higher energy than the corresponding $n^a_+=n^a_-$ state, and
every term in the Hamiltonian, including the fermion mass term,
preserves $n^a_+-n^a_-$.  In other words, states with $n^a_+\not=
n^a_-$ decouple from the Hilbert space constructed on the vacuum
determined below.  The inner product is defined by $\la \Phi_a | \Phi_b
\ra =
\int_{-\infty}^\infty d\wil \int_0^{2\pi} dq^1_+ dq^1_- dq^2_+ dq^2_- ~
\Phi_a^* \,  \Phi_b$.  Note
$\la \Phi_s^{(n,r)}| \Phi_{s'}^{(n',r')} \ra =
 \delta_{ss'}\delta^{rr'} \delta^{nn'}$.

The Hamiltonian has an additional symmetry generated by
homotopically non-trivial large gauge transformations which are given by
$A_\mu \go A_\mu +e^{-1} \d_\mu
\Lambda$, $\psi_a \go e^{-i\Lambda} \psi_a$ where $\Lambda = 2\pi l
x/L$  ($l$ integer).  They preserve the  boundary conditions, inducing
transformations
$\wil \go \wil + 2\pi l$ and  $p^a_\pm \go p^a_\pm - l$.
The $l$=$1$ transformation  is generated by
\beqn
&&U = \exp \Big\{ i\Big( q^1_+ +q^1_- +q^2_+ +q^2_-
             +2\pi \Pi_\Theta \Big) \Big\}  \cr
&&U H U^{-1} =  H  \label{LargeInv}
\eeqn

Under such a gauge transformation
$U \,\Phi_s^{(n,r)} = \Phi_s^{(n+1,r)}$
so that one is naturally led to considering gauge covariant
$\theta$ states
\beqn
\Phi_s^r(\theta) &=& {1\over \sqrt{2\pi}}
\sum_n e^{in\theta} \, \Phi_s^{(n,r)}
\eeqn
which satisfy $\la \Phi_s^r(\theta) | \Phi_{s'}^{r'}(\theta') \ra =
\delta_{ss'} \delta^{rr'} \, \delta_{2\pi}(\theta- \theta')$.  We
shall see below that non-vanishing fermion masses ($m_1,m_2\not= 0$)
absolutely necessitate the $\theta$ states.

There are three effects which the fermion mass term $H_\mass$ brings
about. Firstly, it induces transitions from one $\Phi_s^{(n,r)}$ to
another $\Phi_{s'}^{(n',r')}$. Secondly, it gives a mass $\mu_2$ to
the $\chi_2(x)$ field. Thirdly, it induces interactions among the
$\chi_1$ and $\chi_2$ fields, and zero modes $q^a_\pm$. We restrict
ourselves to the case $m_a \ll \mu$.  The mass $\mu_2$ then depends on
the vacuum structure and must be determined self-consistently.

It is important to realize that in the zero mode sector $H_\mass$
brings about a significant change in the structure of the ground
state.  Although $\Phi_{s=0}^{(n,r)}$ with $r\not=0$ has a higher
eigenvalue of $H_0$, it is just $\pi r^2/L$.  Since $H_\mass$ induces
transitions among various $(n,r)$ states with typical matrix elements
of order $\sim ~mL (\mu L)^{1/2}$, the ground state of the total
Hamiltonian for a large $L$ becomes a superposition of various
$\Phi_0^r(\theta)$ with significant weight for $r\not= 0$.

To see this  more clearly, we first introduce the function
\beqn
&&2 \pi \Delta (x;\mu , L) =  \pi \sum_{n\not= 0}
{ e^{-2\pi inx/L} \over \sqrt{ (2\pi n)^2 + (\mu L)^2 } } \cr
&&\hskip 0.cm = \cases{
 - {1\over 2} \ln 2\Big( 1- \cos{\mybig 2\pi x\over \mybig L} \Big)
   &for $\mu=0$\cr
 K_0(\mu |x|)   - {\mybig \pi\over \mybig \mu L} + 2  \mybig
\int_1^\infty du
\, {\mybig \cosh \mu xu\over \mybig  ( e^{\mu L u}-1 ) \sqrt{u^2-1}}
 &for $\mu>0,  |x| < L$ \cr}
    \label{DeltaFunction}
\eeqn
where $K_0(z)$ is the modified Bessel function.
We shall also encounter the quantity
\beqn
 B(\mu L) &=&
  \exp \Big\{ - 2\pi [\Delta(0; \mu,L) - \Delta(0; 0,L)] \, \Big\} \cr
\noalign{\kern 5pt}
&=& {\mu L\over 4\pi} \exp \bigg\{ \gamma + {\pi\over \mu L}
 - 2 \int_1^\infty  {du \over (e^{ \mu L u} - 1)\sqrt{u^2-1}}  \bigg\}~.
  \label{Bfunction}
\eeqn
In the Schr\"odinger picture we have the following relations for
normal ordering of operators
\beqn
N_\mu[\, e^{i\beta_1 \chi(x)} \, ] \, N_\mu[\, e^{i\beta_2\chi(0)} \,]
 = e^{-\beta_1\beta_2 \Delta(x;\mu,L)} \,
 N_\mu [\, e^{i\beta_1 \chi(x) + i \beta_2 \chi(0) } \, ] \hskip 3.7cm &&\cr
\noalign{\kern 6pt}
N_0 [\, e^{i\beta_1 \chi(x) + i \beta_2 \chi(0) } \, ]
%\cr &&\hskip 0.3cm
= B(\mu L)^{(\beta_1^2 + \beta_2^2) / 4\pi} \,
 e^{ -\beta_1\beta_2 [\Delta(x; \mu,L) - \Delta(x; 0,L)] } \,
N_\mu [\, e^{i\beta_1 \chi(x) + i \beta_2 \chi(0) } \, ]   ~.&&
       \label{useful2}
\eeqn

Now consider the operator
$M_{ab} = \psi^{a\dagger}_- \psi^b_+ = \psibar_a \onehalf (1 + \gamma^5)
\psi_b^{}$  introduced above.
\beqn
&&M_{ab}^S =
\sign(a < b) \cdot
C^{a \dagger}_- C^b_+\cdot e^{ - 2\pi i(p^a_- -  p^b_+)x/L }
\cdot e^{i ( q^a_- + q^b_+ )} \cr
&&\hskip 2cm \times  L^{-1}  ~ N_0[ e^{i\sqrt{2\pi} \chi_1} ]
 ~  N_0[ e^{ i\sqrt{2\pi}(\ep_a \chi_{2-} + \ep_b \chi_{2+})} ]
\eeqn
where $(\ep_1,\ep_2)=(+1,-1)$ and
$\sign (A )  = +$ or $-$,  if $A$ is true or false, respectively  ($S$
stands for Schr\"odinger picture).
In particular, for mass operators $M_{aa}$
\beqn
M_{aa}^S &=& -
C^{a \dagger}_- C^a_+ \cdot e^{ - 2\pi i(p^a_- -  p^a_+)x/L }
\cdot e^{i ( q^a_- + q^a_+ )}
%\cr &&\hskip 1cm \times
\cdot L^{-1} \, \Bbar \,
   N_{\mu_1}[ e^{i\sqrt{2\pi} \chi_1} ]
                 ~  N_{\mu_2}[ e^{i\sqrt{2\pi}  \ep_a \chi_2} ] \\
\Bbar &=&  B(\mu_1 L)^{1/2} B(\mu_2 L)^{1/2} ~.\nonumber
\eeqn
Note $\mu_1 \sim \mu$  for $m/\mu \ll 1$, and we set $\mu_1=\mu$ in
the following. $\mu_2$ is to be determined self-consistently.

It is easy to see that
\beeq
\la \Phi_{s'}^{(n',r')} |
 \left[ \matrix{ M_{11} \cr M_{22} \cr} \right]  | \Phi_s^{(n,r)} \ra
= - L^{-1} \Bbar \, U_{s's} ~
\left[ \matrix{ \delta_{n',n+1} \delta_{r',r-1} \cr
 \delta_{n',n}  \delta_{r',r+1} \cr}  \right]
\eneq
where $U_{s's}= \int_{-\infty}^\infty dx \, u_{s'}(x+\pi) u_s(x)
=(-)^{s'-s} U_{ss'}$
evaluates to
\beqn
&&U_{s's}(\mu L)=e^{-\pi/2\mu L} \sum_{p=0}^s
{(-)^p \sqrt{ s'! s!} \over p! (s'-s+p)! (s-p)!}
\Big( {\pi\over \mu L} \Big)^{(s'-s+2p)/2}
\eeqn
for $s' \ge s$.
Notice that for $\mu L\gg 1$, $U_{s's}\sim \delta_{s's}$ so that we
can safely restrict ourselves to $s=0$ states in constructing the
vacuum. Transitions to $s\ge 1$ states are very small.
For $\mu L\ll 1$ the magnitude of $U_{s's}$ itself is suppressed
exponentially ($\sim e^{-\pi/2\mu L}$), however the ratio of
$U_{s0}$ ($s\ge 1$) to $U_{00}= e^{-\pi/2\mu L}$ becomes large.
Since in this paper we are concerned with the large volume physics,
we can ignore transitions to $s\ge 1$ states, and suppress the index
$s$ henceforth. A more full account incorporating transitions to
higher $s$ states is reserved for future publications.  We remark that
the effect of $H_\mass$ becomes negligibly small for $\mu L \ll 1$,
and so most of the qualitative results below would be insensitive to
transitions to higher $s$ states.

Matrix elements of  $H_\mass$
% =\int \,dx~( m_1 M_{11} + m_2 M_{22} + {\rm h.c.})$
take a simple form in a $\theta$-$\vphi$ basis defined by
\beeq
\Phi(\theta;\vphi)  = {1\over \sqrt{2\pi} } \sum_r e^{ir\vphi} \,
     \Phi^{r}(\theta) ~.
\eneq
They are given by
\beqn
\la  \Phi (\theta';\vphi') | H_\mass | \Phi(\theta;\vphi) \ra
&=& - 2\mbar \Bbar  \, e^{-\pi/2\mu L}
\cos (\vphi+\delbar) ~
 \delta_{2\pi} (\theta-\theta') \delta_{2\pi} (\vphi-\vphi') \cr
m_1e^{-i\theta} + m_2 &=& \mbar (\theta) ~ e^{+ i \delbar(\theta)}
   \hskip 1cm  (\mbar >0)~.  \label{mdeltabar}
\eeqn
In the $SU(2)$ symmetric case ($m_a=m $)
\beqn
&& \delbar(\theta) = - {\theta \over 2}
  + \pi~ {\tt floor}\bigg( {\theta +\pi\over 2\pi}  \bigg) ~~,~~
\mbar(\theta) = 2m \, \big| \cos \onehalf \theta  \big|
\eeqn
where {\tt floor}($x$) is the maximum integer which does not exceed $x$.
Notice that
$\delbar(\theta)$ has discontinuities at $\theta=\pm \pi, \pm 3\pi,
\cdots$ where $\mbar(\theta)$ vanishes.

We write the vacuum in the form
\beeq
| \vPhi (\theta) \ra
= \int_0^{2\pi} d\vphi
    ~ f(\vphi + \delbar) \, | \Phi(\theta;\vphi) \ra  ~.
     \label{vacuumF}
\eneq
Since $\pi r^2/L$ in $H_0$ acts on $\Phi(\theta;\vphi)$ as $-(\pi/L)
(\d^2/\d\vphi^2)$, the eigenvalue equation
$(H_0 + H_\mass - E )~|\vPhi(\theta) \ra =0$ leads to
\beqn
&&\Big( - {d^2\over d\vphi^2} -  \kappa \, \cos \vphi \,\Big)
 f(\vphi) = \ep  \, f(\vphi)  \cr
\noalign{\kern 5pt}
&&\kappa = {2\over \pi} \, \mbar L   \Bbar e^{-\pi/2\mu L}
   \label{mykappa}
\eeqn
where $\ep= EL/\pi$.  This is nothing but the Schr\"odinger
equation in a potential $-\kappa \cos \vphi$ (the quantum pendulum)
whose  ground state satisfies
$f(\vphi)=f(\vphi)^*=f(-\vphi)$. Eq.\ (\ref{mykappa}) is easily solved for
an arbitrary value of $\kappa$ numerically. Thus $f(\vphi)$, and
therefore the structure of the vacuum $\vPhi(\theta)$, is controlled by
$\kappa$.

In two limits $f(\vphi;\kappa)$ can be found analytically.
\beqn
{\rm for~}  \kappa \ll 1, &&
\ep = - {\kappa^2\over 2} + {\rm O}(\kappa^3)  \cr
&&f(\vphi) = %{1\over \sqrt{2\pi}} \, \Big(
   1 + \kappa \cos \vphi -  \kappa^2 \Big( {1\over 4}
   - { \cos 2\vphi\over 8} \Big)
+ {\rm O}(\kappa^3)   \cr
\noalign{\kern 6pt}
{\rm for~}  \kappa \gg 1, &&
\ep ~~ = - \kappa + \sqrt{{\kappa\over 2}} \cr
&&f(\vphi) =  e^{- 2^{-3/2} \kappa^{1/2} \,  \vphi^2  }
  \hskip 1cm (|\vphi|<\pi).  \label{ffunction}
\eeqn
{}From the definition of $\kappa$ in (\ref{mykappa}) we see that
the ~$\mbar\go 0$  and $L\go \infty$ limits
do not commute with each other.

For later convenience we  introduce
\beeq
F_r(u) = \int_0^{2\pi} d\vphi \, f(\vphi+\onehalf u)^*
f(\vphi-\onehalf u) \, e^{ir\vphi} \bigg/
\int_0^{2\pi} d\vphi \, \big| f(\vphi) \big|^2
 \label{Ffunction}
\eneq
which satisfies $F_r(u)^*$=$F_{-r}(u)$=$F_r(-u)$=$F_r(u)$; we
denote $F_r=F_r(0)$ and $F(u)=F_0(u)$,  and recall that $F_r$
depends implicitly on $\kappa$ through $f(\vphi;\kappa)$.
Further we write
 $\la M \ra_\theta=\la \vPhi(\theta) | M| \vPhi(\theta) \ra\Big/\la
\vPhi(\theta) | \vPhi(\theta) \ra $.

A useful formula is
\beqn
&&\la e^{ir_1(q^1_++q^1_-) + ir_2(q^2_++q^2_-) } \,
  e^{-iu(p^1_\pm - p^2_\pm)} \ra_\theta
= \la e^{+iu(p^1_\pm - p^2_\pm)} \,
 e^{ir_1(q^1_++q^1_-) + ir_2(q^2_++q^2_-) }  \ra_\theta \cr
\noalign{\kern 6pt}
&&\hskip 3.cm =e^{+ir_1\bar\delta_1 + ir_2\bar\delta_2} \,
e^{-\pi(r_1+r_2)^2/2\mu L} ~ F_{r_1-r_2}( u) \, e^{i(r_1-r_2)u/2} \cr
&&\hskip .0cm \bar\delta_1(\theta) = -\theta - \bar\delta(\theta) \next
\bar\delta_2(\theta) = + \bar\delta(\theta)
    \label{useful3}
\eeqn
where $r_1$ and $r_2$ are integers.
Then expectation values of $M_{ab}$'s are
\beqn
&&\la M_{ab} \ra_\theta
 = -\delta_{ab} ~ e^{i\delbar_a(\theta)}
\,L^{-1}  \Bbar  e^{- \pi/2\mu L}   \, F_1      \label{expect1}
\eeqn
{}From (\ref{ffunction})
\beeq
F_1 =
\cases{\kappa &for $\kappa \ll 1$\cr
      \exp \big\{ -(2\sqrt{2\kappa})^{-1} \big\}
       \sim 1  &for $\kappa \gg 1$.\cr}  \label{F1behavior}
\eneq
The behavior of $F_1$  with $\kappa$ is depicted in Fig.\ 1.

\begin{figure}[htb]
\epsfxsize= 14cm
\epsffile[40 100 550 550]{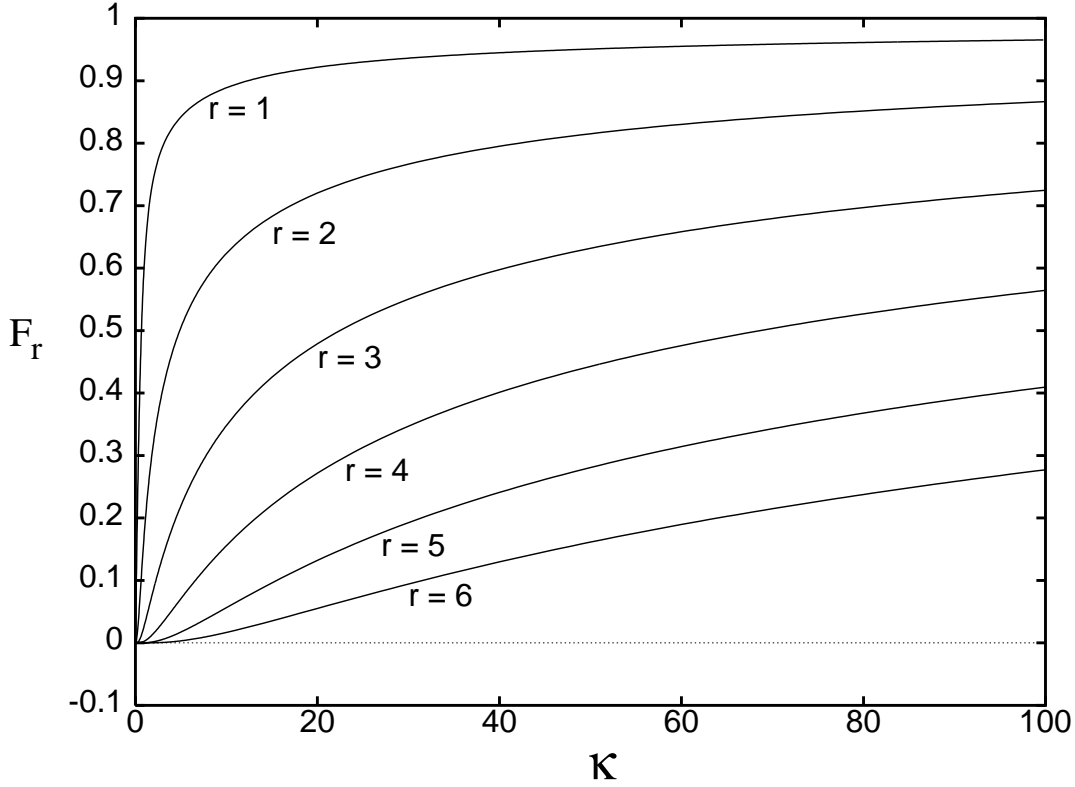}
\vskip -1cm
\caption{$F_r=F_r(0)$ defined in (30)
are plotted as functions of $\kappa$.  Their behavior at $\kappa \ll 1$
and $\kappa \gg 1$ is given by (33) and (45). }
\label{fig:1}
\vskip 0.5cm
\end{figure}

We can now determine $\mu_2$.  Taking the vacuum expectation value
in the zero-mode sector, one finds
\beqn
&&\la H_\mass\ra_{\rm zero~modes} \cr
&&=  -\int dx \, {\Bbar\over L} \, e^{-\pi/2\mu L} \,F_1
\sum_{a=1}^2
\Big\{ ~m_a \,   e^{i\delbar_a} \, N_\mu[e^{i\sqrt{2\pi} \chi_1} ]
N_{\mu_2}[e^{i\ep_a\sqrt{2\pi} \chi_2} ]    + ~ {\rm h.c.} ~\Big\} \cr
&&=\int dx \, {\Bbar\over L} \, e^{-\pi/2\mu L} \,F_1
{}~ \Big\{ - i\sqrt{2\pi}
  (\sum_a m_a \,e^{i\delbar_a}  - {\rm h.c.} ) \chi_1 \cr
&&\hskip 1.5cm - 2\pi  (\sum_a \ep_a m_a \,e^{i\delbar_a}
 + {\rm h.c.} ) \chi_1 \chi_2
 + 2 \pi \mbar (\chi_1^2 +\chi_2^2 ) + {\rm O}(\chi^3)
\Big\}
\eeqn
In the $SU(2)$ symmetric case $m_a\,   e^{i\bar\delta_a }=
m e^{-i\theta/2}(-1)^{{\tt floor}[(\theta+\pi)/2\pi]}$
and the $\chi_1 \chi_2$ term vanishes.
The $\chi_1$ term is proportional to
$m\sin\onehalf\theta$ so that a small shift of $\chi_1$ field is
necessary for $\theta\not= 0$, however the correction to $\mu_1$ is minor.
$\mu_2$ is determined by
\beeq
\mu_2^2 = {4\pi\over L} \, \mbar \Bbar\,  e^{-\pi/ 2\mu L}
\,F_1  = {2\pi^2\over L^2} \, \kappa \, F_1 ~~.
            \label{massEquation}
\eneq
$F_1$ is a function of $\kappa$, and
$\kappa$  depends on $\mu_2$.  Employing
 (\ref{Bfunction}), (\ref{mykappa}), and (\ref{F1behavior}), and
classifying cases according to whether $\kappa$, $\mu L$, $\mu_2L$ are
 large or small
%$\gg 1$ or $\ll 1$,
one finds
\beeq
\mu_2 = \cases{
2\sqrt{2}\,  \mbar \, e^{-\pi/2\mu L}
     &for  $1 \gg \mu L \gg \mu_2 L$\cr
2\sqrt{2} \, \mbar\,\Big({\mybig \mu L e^\gamma\over\mybig 4\pi} \Big)^{1/2}
     &for $\mu L \gg 1 \gg \mu_2 L ,  ~~\mbar L (\mu L)^{1/2} \ll 1$\cr
(e^{2\gamma} \, \mbar^2 \mu )^{1/3}
     &for $\mu L \gg \mu_2 L \gg 1$, ~~$\mbar L (\mu L)^{1/2} \gg 1$\cr}
\eneq
which reproduces  Coleman's result \cite{Coleman} in Minkowski spacetime
 when $\mbar L (\mu L)^{1/2} \gg 1$.
However, if the $\mbar\go 0$ limit is taken with fixed $L$, then
$\mu_2={\rm O}(\mbar)$, i.e.\ $\mu_2$ as a function of $m$ and $\theta$
has a smooth limit at $m\cos\onehalf\theta=0$.

Furthermore, combining (\ref{expect1}) and (\ref{massEquation}), one
finds
\beeq
\la M_{ab} \ra_\theta =
- \delta_{ab} ~ e^{i\delbar_a(\theta)}
{\mu_2^2\over 4\pi\mbar}~.  \label{expect2}
\eneq
It follows that in the $SU(2)$ symmetric case
\beeq
\la \psibar_a\psi_a \ra_\theta = \cases{
- {\mybig 8\over \mybig\pi} \, m  \, e^{-\pi/\mu L} \, \cos^2 \onehalf
\theta
  &for  $\mu L \ll 1$\cr
-{\mybig 2e^\gamma\over\mybig \pi^2} \, m\mu L \, \cos^2 \onehalf \theta
    &for $\mu L \gg 1 \gg
     m L (\mu L)^{1/2} \, | \cos \onehalf \theta |$\cr
-\Big( {\mybig  e^{4\gamma}\over \mybig 4\pi^3} \, m \mu^2
 \, \cos^4 \onehalf \theta  \Big)^{1/3}
  &for $m L (\mu L)^{1/2}\, | \cos \onehalf \theta | \gg 1$\cr}
\eneq
  which agrees with
Smilga's estimate of the condensate in Minkowski spacetime \cite{Smilga}
when $m L (\mu L)^{1/2}\, | \cos \onehalf \theta | \gg 1$.
The singlet four-fermi operator
$M_0=\psi^{1\dagger}_-\psi^{2\dagger}_- \psi^2_+\psi^1_+$ is
\beeq
 M_0(x)  = e^{i(q^1_-+q^1_++q^2_-+q^2_+)}
{1\over L^2}B(\mu L)^2  N_\mu[ e^{i\sqrt{8\pi}\, \chi_1(x)} ]
\eneq
apart from irrelevant operators which act as an identity operator on
physical states.  Hence
\beqn
\la M_0(x) \ra_\theta
&=&e^{-i\theta} \, {1\over L^2} B(\mu L)^2 \, e^{-2\pi/\mu L}    \cr
\noalign{\kern 8pt}
&\sim &
 e^{-i\theta} \Big( {\mu e^\gamma\over 4\pi} \Big)^2 \hskip .5cm
{\rm for~} \mu L \gg 1  \label{singlet}
\eeqn
which is insensitive to the values of $m_a \ll \mu$.  It is
non-vanishing in the massless, infinite volume limit;
the associated chiral $U(1)$ symmetry is broken by the anomaly.
We note that  the corresponding quantity in the $N$=1 case behaves similarly
\cite{HH,Sachs}:   $\la \psibar\psi \ra_\theta
=-e^{-i\theta} L^{-1} B(\mu L) e^{-\pi/\mu L}$ where
$\mu = \sqrt{e^2/\pi}$.

We stress that from the definition (\ref{mdeltabar}), $\mbar(\theta)$
and $\delbar(\theta)$ are periodic in $\theta$ with period $2\pi$.
Therefore condensates (\ref{expect2}) and (\ref{singlet}) also are
periodic functions of $\theta$ with period $2\pi$.  The phase
$\delbar(\theta)$ has a discontinuity at $\theta=\pm \pi$    where $\mu_2$
and  $\la M_{aa}\ra_\theta$ vanish.   In this regard it is important to
recognize the non-trivial  dependence of the vacuum wave function
$f(\vphi)$ on $\theta$ through $\kappa$.

When one of fermion masses, $m_a$, vanishes, the Hamiltonian has an
additional symmetry generated by the conserved gauge variant   chiral
charge
\beeq
\tilde Q^5_{aa} = \int_0^L dx \,
\tilde j^{50}_{aa} = p^a_+ + p^a_-  \next [H, \tilde Q^5_{aa} ] =0 ~.
\eneq
Consider the two cases: (a) $m_1\go 0, m_2=m$ ~~and~~ (b)  $m_1=m , m_2\go 0$.
$\delbar=0$ and $-\theta$ in cases (a) and (b), respectively.
In both cases $\mbar=m$.  Further,
$e^{i\beta \tilde Q^5_{aa} } \, | \Phi(\theta,\vphi) \ra
= e^{2i\alpha\beta} \,
 | \Phi(\theta+2\beta,\vphi+\delta_{a2}\cdot 2\beta) \ra$.
Combining these with (\ref{vacuumF}), one finds that
\beeq
e^{i\beta \tilde Q } \, |\vPhi(\theta) \ra =
 e^{2i\alpha\beta} \, | \vPhi(\theta+ 2\beta) \ra
\eneq
where $\tilde Q= \tilde Q^5_{11}$ or $\tilde Q^5_{22}$ in cases (a) or
(b), respectively.   This establishes the equivalence of all $\theta$
vacua. In particular, $E(\theta)$ is $\theta$-independent.

It is straightforward to evaluate various correlation functions.
For correlators of $M_{ab}$ with $a\not= b$ we need, in addition to
(\ref{useful2}) and (\ref{useful3}),
\beqn
\la N_0[\, e^{i\alpha\chi_{2+} (x) - i\alpha\chi_{2-} (x)
  + i\beta\chi_{2+}(y) - i\beta\chi_{2-} (y) } \,]\ra  \hskip 6.6cm && \cr
\noalign{\kern 3pt}
\hskip 0.0cm = B(\mu_2 L)^{(\alpha^2+\beta^2)/4\pi} \,
e^{-\alpha\beta \{ \Delta(x-y;\mu_2,L) - \Delta(x-y;0,L) \} }
\, e^{- \onehalf (\alpha^2+\beta^2)  h(0;\mu_2,L)
       - \alpha\beta h(x-y; \mu_2 ,L) } && \cr
\noalign{\kern 8pt}
\hskip 0cm   h(x;\mu,L) = {1\over 2L} \sum_{n\not= 0}
{ \mu^2 \over \omega_n(0)^2 \omega_n(\mu) } ~ e^{i p_n x} \hskip 4.5cm  &&
     \label{useful4}
\eeqn
Then, to leading order in $m/\mu$,
\beqn
%\left[ \matrix{
\la M_{aa}(x) M_{aa}(0) \ra_\theta \hskip .55cm
&=&\phantom{-}  D(x; -,-) ~ e^{2i\delbar_a} ~
   e^{-2\pi/\mu L} \, F_2\cr
\la M_{aa}(x) M_{bb}(0) \ra_{\theta, a\not= b}
&=& \phantom{-} D(x; -,+) ~ e^{-i\theta} ~  e^{-2\pi/\mu L} \cr
\la M_{ab}(x) M_{ba}(0) \ra_{\theta, a\not= b}
&=& -  D(x; -,+) ~ e^{-i\theta} ~  e^{-2\pi/\mu L}
   ~F\Big({2\pi x\over L} \Big) \,
    e^{- 2\pi \{ h(0;\mu_2, L) - h(x;\mu_2, L) \} }  \cr
\la M_{aa}^\dagger(x) M_{aa}(0) \ra_\theta \hskip .55cm
&=& \phantom{-} D(x; +,+) \cr
\la M_{ab}^\dagger(x) M_{ab}(0) \ra_{\theta, a\not= b}
&=&\phantom{-} D(x; +,+)  ~~~F\Big({2\pi x\over L} \Big) \,
    e^{- 2\pi \{ h(0;\mu_2, L) - h(x;\mu_2, L) \} } \cr
\la M_{aa}^\dagger(x) M_{bb}(0) \ra_{\theta,a\not= b}
&=& \phantom{-}D(x; +,-) ~e^{i(\delbar_b -\delbar_a)}~ F_2 \cr
\noalign{\kern 5pt}
{\rm where~~} D(x; \sigma_1,\sigma_2)
&=& ~ {1\over L^2} \, B(\mu L) B(\mu_2 L)
{}~ e^{2\pi \sigma_1\Delta(x;\mu,L) + 2\pi \sigma_2 \Delta(x;\mu_2,L)} ~.
   \label{correlator}
\eeqn
All other correlators vanish.
$F_2$ as a function of $\kappa$ is depicted in Fig.\ 1.  In particular
\beeq
F_2 =
\cases{{3\over 8}\kappa^2 &for $\kappa \ll 1$\cr
      \exp \Big\{ -\sqrt{2/\kappa} \, \Big\}
       \sim 1  &for $\kappa \gg 1$.\cr}  \label{F2behavior}
\eneq
Note that except for the dominant
correlators  $\la M_{aa} M_{bb}\ra_{\theta, a\not= b}$ and
$\la M_{aa}^\dagger M_{aa} \ra_\theta$ all others in
(\ref{correlator}) depend on the vacuum wave function $f(\vphi)$
and therefore the parameter $\kappa$.

The behavior at large distances depends on $\mu_2$.  Recall that
\beqn
2\pi \Delta (x;0 , L) &\sim&  -  \ln {2\pi x\over L} \hskip 2.4cm
\hbox{for~~}  L \gg x  \cr
\noalign{\kern 8pt}
2\pi \Delta(x ; \mu ,L)
&\sim& - {\pi\over \mu L} + \sqrt{ {\pi\over 2\mu x}} \, e^{-\mu x}
\hskip 0.5cm {\rm for~~} \mu L \gg \mu x \gg 1  \cr
\noalign{\kern 6pt}
&\sim& - \ln \Big( {\mu x\over 2} e^\gamma \Big)
\hskip 1.75cm {\rm for~~}  \mu L \gg 1\gg \mu x
   \label{DeltaBehavior}
\eeqn
Hence when $\mu_2=0$, i.e.\ $m\cos\onehalf\theta=0$,
\beeq
\left[ \matrix{\la M_{aa}(x) M_{bb}(0) \ra_{\theta, a\not= b}\cr
\la M_{aa}^\dagger(x) M_{aa}(0) \ra_\theta ~~  \cr} \right]
= \left[ \matrix{ e^{-i\theta} \cr 1\cr} \right]
    { \mu e^\gamma\over 8\pi^2 x}
\qquad {\rm for~} \mu L \gg  \mu x \gg 1 ~.
  \label{correlator2}
\eneq
Hence the critical exponent is 1 in accordance with Affleck's result
on $R^1$.    On the other hand, when $\mu \gg \mu_2>0$,
$\la M_{aa} \ra_\theta \not= 0$ and the correlators decay exponentially:
\beqn
&&\left[ \matrix{\la M_{aa}(x) M_{bb}(0) \ra_{\theta}
   -\la M_{aa} \ra_{\theta}\la M_{bb}\ra_{\theta} \cr
\la M_{aa}^\dagger(x) M_{bb}(0) \ra_\theta
 - \la M_{aa}^\dagger\ra_\theta \la  M_{bb} \ra_\theta\cr} \right] \cr
\noalign{\kern 10pt}
&&\hskip .5cm \sim \mp e^{\pm i\delbar_a + i\delbar_b} \,
{\mu \mu_2 e^{2\gamma}\over  16\pi^2}
\Bigg\{~
\sqrt{{\pi\over 2\mu x}} \, e^{-\mu x}
  + \ep_a\ep_b \sqrt{{\pi\over 2\mu_2 x}} \, e^{-\mu_2 x} ~\Bigg\}  \cr
\noalign{\kern 10pt}
&&\hskip 5.cm  {\rm for~}   \mu L \gg\mu_2 L \gg  \mu_2 x \gg 1~.
    \label{correlator3}
\eeqn
At short distances
\beqn
&&\la M_{aa}(x) M_{bb}(0) \ra_{\theta, a\not= b}
= e^{-i\theta} \Big({\mybig \mu  e^{\gamma}\over \mybig
              4\pi}\Big)^2\cr
&&
\la M_{aa}^\dagger(x) M_{aa}(0) \ra_\theta ~~
= ~~ {\mybig 1 \over \mybig 4 \pi^2 x^2}  \\
\noalign{\kern 5pt}
&&\hskip 0.cm {\rm for}~~
\mu L \gg 1  \gg \mu x ~,~ \mu_2=0 ~ ~{\rm or}~~
%\cr  &&\hskip 2.cm
\mu L \gg \mu_2 L \gg 1 \gg  \mu x \gg \mu_2 x    \nonumber
     \label{correlator4}
\eeqn
For the chiral $U(1)$ condensate (\ref{singlet}),
\beqn
&&\la M_0^\dagger(x) M_0(0)\ra_\theta
= {1\over L^4} \,B(\mu L)^4 \,  e^{+ 8\pi\Delta(x;\mu,L)}  \cr
\noalign{\kern 8pt}
&&\hskip 1cm =\cases{
\Big( {\mybig \mu e^\gamma\over \mybig 4\pi} \Big)^4
   \exp \Big( \sqrt{ {\mbig 8\pi\over \mybig \mu x} } \, e^{-\mu x}
\Big)
     &for $\mu x \gg 1, \mu L \gg 1,  x/ L \ll 1 $\cr
{\mybig 1\over \mybig ( 2\pi x)^4 }
 &for $\mu x \ll 1, \mu L \gg 1,  x/ L \ll 1 $\cr}
\eeqn
The correlator $\la M_0^\dagger(x) M_0(0)\ra_\theta -
|\la M_0 \ra_\theta |^2$ shows an exponential fall-off.

An interesting extension of this work comes from the fact that
quantum spin $\onehalf$ antiferromagnet chains are almost equivalent
to an $N=2$ massless Schwinger model at strong coupling. Our result
supports this correspondence, where staggered magnetization
corresponds to the chiral condensate. The critical exponent is
$\eta=1$, as pointed out by Haldane
\cite{Haldane} and Affleck
\cite{Affleck2} who converted spin chains to nonlinear sigma
models and found that there is no long-range-order in the infinite
volume limit. Our $\chi_2$ field corresponds to the gapless (spinon)
mode in spin chains\cite{Lowenstein2}. We also stress that our result
is valid for arbitrary $L$, and thus should be very useful for
investigating finite size effects in spin chains.

To conclude, we have analyzed the two-flavor massive Schwinger model
on a circle.  We have reduced the system to a one-dimensional quantum
mechanics problem defined by (\ref{mykappa}).  The approximation is
justified for $\mu L$$\gg$1 but with an arbitrary $\mbar L$.  For $\mu
L$$<$1 excitations to higher $s$ states need to be taken into account,
and we expect the system to reduce to a two-dimensional quantum
mechanics problem. Recently Shifman and Smilga \cite{Shifman} have
proposed that with twisted boundary condition for fermions,
excitations with fractional topological number (fractons) become
essential. Our analysis  should straightforwardly generalize to
cases with $N$ flavors of fermions with arbitrary boundary conditions.
We plan to return to these points in separate publications.

\vskip 1cm
\leftline{\bf Acknowledgments}

This work was supported in part by by the U.S.\ Department of Energy
under contracts DE-AC02-83ER-40105 (Y.H.) and by DE-FG02-85ER-40213
(J.H.).   The authors would like to thank H.\ Joos, M.\
Shifman, C.\ Dasgupta, and R.\ Laughlin for many enlightening
discussions.

\vskip 1.cm

\def\ap {{\it Ann.\ Phys.\ (N.Y.)} }
\def\cmp {{\it Comm.\ Math.\ Phys.} }
\def\ijmpA {{\it Int.\ J.\ Mod.\ Phys.} { \bf A}}
\def\ijmpB {{\it Int.\ J.\ Mod.\ Phys.} { \bf B}}
\def\jmp {{\it  J.\ Math.\ Phys.} }
\def\mplA {{\it Mod.\ Phys.\ Lett.} { \bf A}}
\def\mplB {{\it Mod.\ Phys.\ Lett.} { \bf B}}
\def\plB {{\it Phys.\ Lett.} { \bf B}}
\def\plA {{\it Phys.\ Lett.} { \bf A}}
\def\nc {{\it Nuovo Cimento} }
\def\npB {{\it Nucl.\ Phys.} { \bf B}}
\def\pr {{\it Phys.\ Rev.} }
\def\prl {{\it Phys.\ Rev.\ Lett.} }
\def\prB {{\it Phys.\ Rev.} { \bf B}}
\def\prD {{\it Phys.\ Rev.} { \bf D}}
\def\prp {{\it Phys.\ Report} }
\def\ptp {{\it Prog.\ Theoret.\ Phys.} }
\def\rmp {{\it Rev.\ Mod.\ Phys.} }
\def\hep {{\tt hep-th/}}

\end{document}